\documentclass{article}
\usepackage{spconf, amsmath, graphicx, tabularx, array, comment, hyperref}
\usepackage{xcolor}
\usepackage{mwe} 
\usepackage[normalem]{ulem}

\newcommand{\FHC}{{\textsc{FHC}}}
\newcommand{\AFHC}{A\FHC}
\newcommand{\GGT}{GGT}
\newcommand{\SGT}{SGT}

\newcommand{\hospital}{Alder Hey Children's Hospital in Liverpool, England}
\newcommand{\trust}{Alder Hey Children’s NHS Foundation Trust}

\title{Infant hip screening using multi-class ultrasound scan segmentation}

\name{
Andrew Stamper $^{1}$ \quad 
Abhinav Singh $^{1, 2}$ \quad 
James McCouat $^{1, 2}$  \quad
Irina Voiculescu $^{1}$
}
\address{
$^{1}$ Department of Computer Science, University of Oxford, UK \\ $^{2}$ NDORMS, University of Oxford, UK.
}
\begin{document}
%
\maketitle
\begin{abstract}
Developmental dysplasia of the hip (DDH) is a condition in infants where the femoral head is incorrectly located in the hip joint. We propose a deep learning algorithm for segmenting key structures within ultrasound images, employing this to calculate Femoral Head Coverage (\FHC) and provide a screening diagnosis for DDH. To our knowledge, this is the first study to automate FHC calculation for DDH screening. Our algorithm outperforms the international state of the art,  agreeing with expert clinicians on 89.8\% of our test images.

\end{abstract}

\section{Introduction}

\label{sec:intro}
Developmental dysplasia of the hip (DDH) is a common cause of childhood disability, with missed cases representing the largest cause of premature arthritis in young adults~\cite{osteoarthritis}. In newborns, the condition ranges from outward migration of the femoral head to complete dislocation from the socket. DDH diagnosis under three months enables non-invasive treatment; whereas late diagnosis necessitates expensive complex surgery with significant interruption to social development in infancy~\cite{osteoarthritis, DDHarticle2}. Approximately one third of hip replacements in patients under the age of 40 are due to DDH~\cite{osteoarthritis}. 

Employing experts to perform and report scans routinely is a significant cost barrier to scanning {\em all\/} newborns. Therefore, only those with risk factors undergo an ultrasound.

\begin{figure}[t]
\centering
\begin{minipage}[b]{.36\linewidth}
  \centerline{\includegraphics[height=\linewidth, trim={0.5cm 3cm 3cm 0.5cm}, clip]{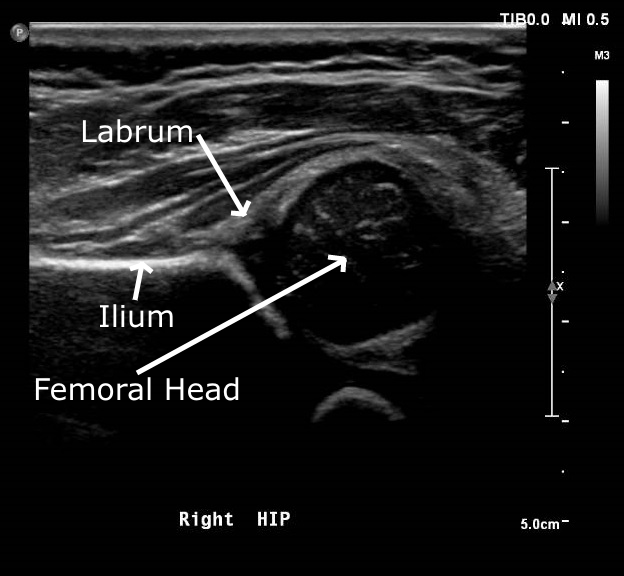}}
  \centerline{(a) Annotated hip joint}\medskip
\end{minipage}
~~~~~
\begin{minipage}[b]{0.36\linewidth}
  \centerline{\includegraphics[height=\linewidth, trim={0.5cm 3cm 3cm 0.5cm}, clip]{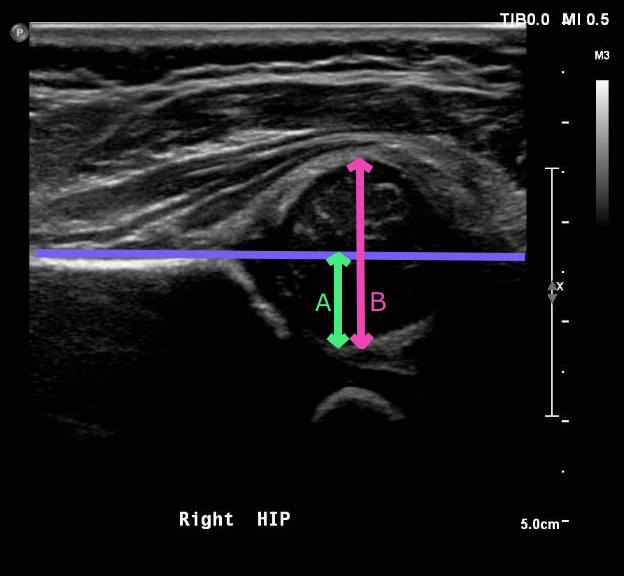}}
  \centerline{(b) \FHC\ Measurements}\medskip
\end{minipage}
\caption{(a) Static ultrasound of hip joint, annotated with femoral head, ilium and labrum. 
(b)~Measurements required for \FHC\ evaluation: $\FHC\% = \frac{A}{B}{\times}100$.
}
\label{fig:SegmentationANDFHC}
\end{figure}

Ultrasound imaging is more sensitive (clinically) than a physical examination and is ideal for screening programmes as it is safe and portable \cite{dogruel2008clinical}. The Femoral Head Coverage~(\FHC) method is used to provide objective evidence where a straight line is drawn extending along the upper edge of the ilium (Fig.~\ref{fig:SegmentationANDFHC}) and the femoral head proportion on each side of the line is measured \cite{morin1985infant}. Clinically, a decision value of $50\%$ coverage is used, with a larger percentage below the line (\FHC$>$50\%) identified as healthy and a larger percentage above the line (\FHC$\leq$50\%) termed DDH \cite{gunay2009correlation}.

State of the art literature for automatic DDH evaluation uses the Graf method \cite{relatedworks1, relatedworks2, relatedworks3}. However, this method has extremely high inter-operator variability and it only achieves approximately 85\% agreement with clinicians.
Our reproducible algorithm takes automated numerical measures of the \FHC\ (\AFHC) to classify the presence or absence of DDH. It does so by segmenting key anatomical structures from static 2D ultrasound images: ilium, femoral head and labrum (Fig.~\ref{fig:SegmentationANDFHC}). Clinicians are able to identify these structures and diagnose by simply inspecting the scan (Gestalt laws)~\cite{cianci2015gestalt}.




\begin{figure}[t]
\centering
\begin{minipage}[b]{.3\linewidth}
  \centerline{\includegraphics[width=\linewidth]{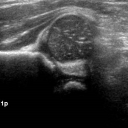}}
  \centerline{(a)}
\end{minipage}
%
%
~~~~
\begin{minipage}[b]{0.3\linewidth}
  \centerline{\includegraphics[width=\linewidth]{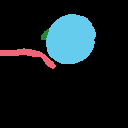}}
  \centerline{(b)}
\end{minipage}

\begin{minipage}[b]{.3\linewidth}
  \centerline{\includegraphics[width=\linewidth]{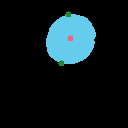}}
  \centerline{(c)}\medskip
\end{minipage}
\hfill
\begin{minipage}[b]{0.3\linewidth}
  \centerline{\includegraphics[width=\linewidth]{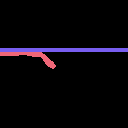}}
  \centerline{(d)}\medskip
\end{minipage}
\hfill
\begin{minipage}[b]{0.3\linewidth}
  \centerline{\includegraphics[width=\linewidth]{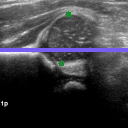}}
  \centerline{(e)}\medskip
\end{minipage}

\caption{Generating \AFHC. (a)~Raw image, (b)~Automated segmentation, (c)~Femoral head centroid~(red) and extrema~(green), (d)~Ilium line (purple), (e)~Key points for \AFHC. Inferred is the diagnosis that DDH is present.}
\label{fig:process}
\end{figure}

In this study, two experienced clinicians have generated the one-bit ground truth determining the presence or absence of DDH as perceived through their clinical experience. Hereafter, this clinical decision is referred to as the Gestalt Ground Truth (\GGT). Separately, a third clinician has segmented out anatomical features and has taken \FHC\ measurements, referred to as the Segmented Ground Truth (\SGT). 




\section{Method}
\label{sec:method}
Our algorithm first uses a multi-class convolutional neural network to segment the key anatomical structures mentioned earlier (Fig.~\ref{fig:process}b). The presence of these structures is a marker that the correct ultrasound image has been acquired.
%
%
To measure \AFHC, the algorithm locates the extrema on the femoral head, and the straight line extending from the uppermost edge of the ilium. 
%
%
Our algorithm locates the upper and lowermost points of the femoral head by first locating the centroid of the head mask and then flooding the mask (Fig.~\ref{fig:process}c).

To identify the ilium line (Figure \ref{fig:process}d) we first isolate its horizontal part. We identify the topmost pixel of the ilium mask $(x,y)$ s.t.\ $(x,y){\in}\mbox{mask}$ but $(x,y{+}1){\notin}\mbox{mask}$, then partition the boundary into short segments. 
A line of best fit is drawn through each segment, and the angles between adjacent lines are evaluated. The point at which the ilium turns (downwards) from horizontal is therefore the point at which this angle is maximal. Pixels to the left of this point are identified as part of the horizontal upper ilium edge and a horizontal line is drawn through their average height (Fig.~\ref{fig:process}e)

Using the two extrema of the femoral head, and the produced ilium line (Fig.~\ref{fig:process}e) determine \AFHC, and thus DDH diagnosis is derived. A decision value of \FHC$\leq$50\% is used to provide a screening diagnosis for DDH.

\section{Dataset}
\label{sec:dataset}

The data has been obtained directly from static ultrasounds taken at the \hospital\ as part of routine screening.
All images were taken at under 12 weeks and each subject had either an abnormal clinical examination for DDH, or a positive risk factor for DDH~\cite{GovPublicationOnNipe}.

To generate the \GGT, two experts classified the images into three groups: Normal (centred), and two severities of DDH: Dysplastic or Dislocated (requires monitoring and treatment). A third clinician  derived the \SGT\ by segmenting three key structures from each ultrasound scan: the femoral head, ilium and labrum (Fig.~\ref{fig:process}b). These masks were verified by the two experts.

	
The dataset contains 94 right and 96 left hip scans.  The left hip scans are reflected to produce 190 right-hip-like scans. 
Data was split into testing, validation and training by a 50:25:25 \% split. To ensure a fair representation of each type of scan in each of these sets, the dataset was first split into normal (71 images, 37\%), dysplastic (66 images, 35\% ) and dislocated collections (53 images, 28\%) (according to \GGT), and these collections were combined proportionally to create balanced training, validation and testing sets.

The raw images were of non uniform width and height. Each image was pre-processed, cropping to $384{\times}384$ pixels, then down sampled (via max pooling) to $128{\times}128$ pixels.

\begin{figure}[t]

  \centerline{\includegraphics[width=0.9\linewidth, trim={6cm 5cm 6cm 2cm}, clip]{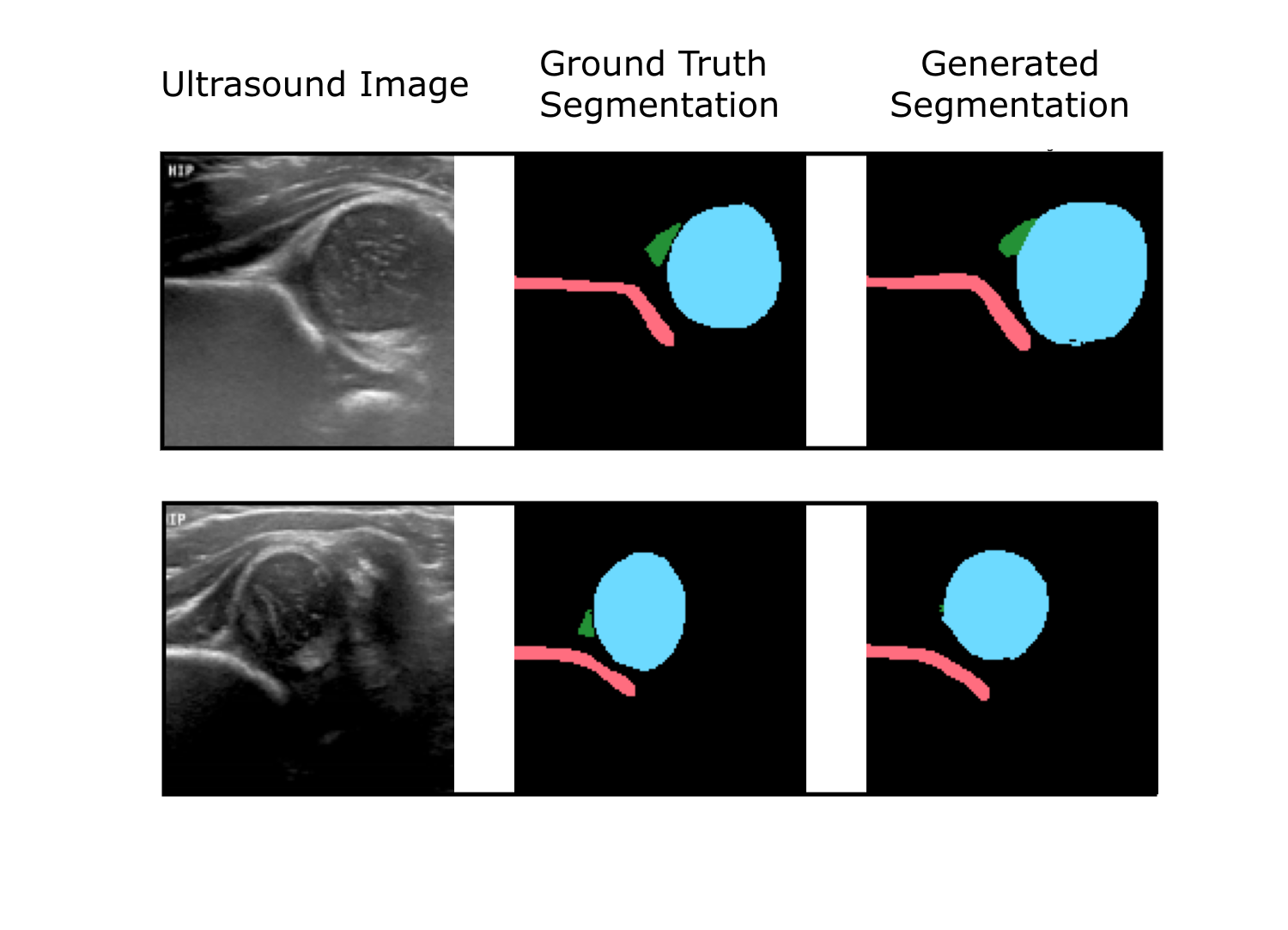}}
  

\caption{Qualitative display of automated segmentation results.}
\label{fig:qualitativeAnalysis}
\end{figure}

\section{Experiments}
\label{sec:experiments}


A modified multi-class U-Net CNN is used to segment the key anatomical structures.
This U-Net is a variant of that introduced in~\cite{UNET}, modified to use $128{\times}128$ input and output, `SAME' padding, a single convolution per encoder and decoder layer (rather than two in the original design), down sampling to a resolution of only $4{\times}4$ pixels, and a batch normalisation layer before each ReLu function. These modifications reduce the model's tendency to overfit, a key consideration given the restrictive size of our training set. A batch size of 4 and an ADAM optimizer (learning rate: 0.001) are used with the cross entropy loss function. 
Data augmentation reduces further the risk of over-fitting, and is conducted in a random order, with the degree of each modification determined randomly within the following ranges. Overall image brightness: $0.5$--$1.5$, Gaussian blur: sigma $0.0$--$2.0$, $x$-axis or  $y$-axis scaling: $0.9$--$1.1$, $x$-axis or $y$-axis translation: $0$--$10$\%, rotation: ${-15}^{\circ}$ to $15^{\circ}$.


The model trained for 50 epochs.
The reduced number of parameters makes it particularly lightweight, taking less than 15 seconds per epoch to train on a low power laptop CPU: Intel i7-8550U@1.80GHz. 

\begin{figure}[t]
  \centering
  \centerline{\includegraphics[width=0.8\linewidth]{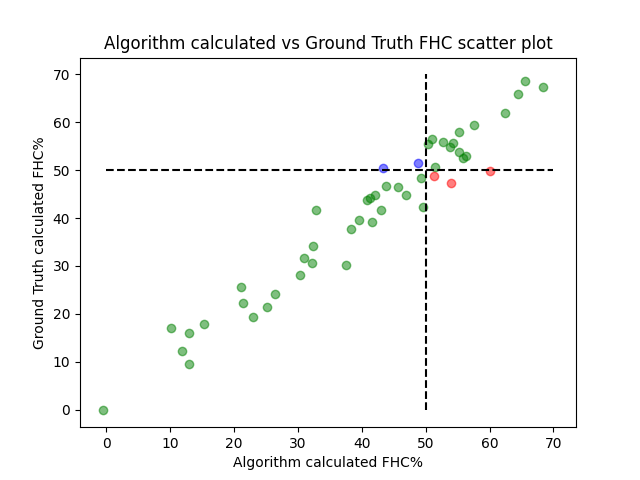}}
\caption{Scatter plot of \AFHC\ diagnosis versus \SGT. The 50\% decision value is shown in black. Patients with correct diagnosis in green, false positives in blue, false negatives in red.}
\label{fig:scatterFHC}
\end{figure}

\section{Results and Discussion}
\label{sec:results_and_discussion}

The segmentation model identifies all three anatomical structures in all scans in the test set (Tab.~\ref{tab:segmentationresults}). The femoral head has good overlap scores (Dice, Precision, Recall, Sensitivity, Specificity, etc.) but, as expected from the presentation of the data, less accurate boundaries (Hausdorff and Average Symmetric Surface Distance). The ilium is well identified in both overlap and Average Symmetric Surface Distance scores but performs less impressively in Hausdorff Distance, implying the majority of boundary pixels are well identified but the odd pixel in the boundary may be displaced (since HD is the maximal boundary displacement across the whole image). While overlap scores of the labrum appear somewhat unimpressive, 
they serve well the purpose of its identification.

Fig.~\ref{fig:qualitativeAnalysis} compares the \SGT\ and automated segmentation of two sample images: one typical performance and one of its worst segmentations (identifying the labrum by only five pixels). 
Having compared the segmented masks against \SGT\ segmentation masks, we focus on our primary interest in the system's diagnostic performance. Comparing the diagnosis inferred from \AFHC\ against \SGT\  we obtain 89.8\%  accuracy, 90.6\%  sensitivity and 88.2\% specificity  (Tab.~\ref{tab:diagnosis}).
%
%
%
The scatter plot in Fig.~\ref{fig:scatterFHC} illustrates the strong correlation between the automated \AFHC-based diagnosis and \SGT.

To compare the inter-rater reliability we use Cohen's Kappa coefficient $\kappa$~\cite{Marston2009IntroductorySF}. The agreement between the clinicians' \GGT\ and \SGT\ is 85.7\% on the test data, with $\kappa{=}0.689$. By contrast, the \AFHC-based diagnosis agrees with \SGT\ on 89.8\% of the test data, with $\kappa{=}0.778$. The three pairwise comparisons are summarised in Tab.~\ref{tab:FHCagreement}.
%
The agreement between all three diagnoses amounts to 81.6\% of the test set, measured with a Fleiss’ Kappa of 0.734.


\begin{table}[ht]
		\begin{center}
			\begin{tabular}{ | c | c | c | c |}
				\hline
Metric & Ilium & Femoral Head & Labrum \\ \hline 
\hline 
DSC & 0.857$\pm$0.049 & 0.924$\pm$0.03 & 0.71$\pm$0.156 \\ \hline 
TPVF & 0.889$\pm$0.058 & 0.982$\pm$0.022 & 0.727$\pm$0.194 \\ \hline 
TNVF & 0.996$\pm$0.002 & 0.979$\pm$0.011 & 0.998$\pm$0.001 \\ \hline 
FPVF & 0.004$\pm$0.002 & 0.021$\pm$0.011 & 0.002$\pm$0.001 \\ \hline 
FNVF & 0.111$\pm$0.058 & 0.018$\pm$0.022 & 0.273$\pm$0.194 \\ \hline 
Prec & 0.833$\pm$0.076 & 0.875$\pm$0.058 & 0.723$\pm$0.141 \\ \hline 
\hline
HD & 5.784$\pm$12.829 & 6.081$\pm$7.774 & 4.29$\pm$2.484 \\ \hline ASSD & 0.718$\pm$0.367 & 1.888$\pm$1.057 & 1.103$\pm$0.881 \\ \hline 
			\end{tabular}
		\end{center}

		\caption{Mean and standard deviation of overlap measures over the test set images. Dice Similarity Coefficient (DSC), True Positive, True Negative, False Positive, False Negative Volume Fractions (TPVF, TNVF, FPVF and FNVF) and Precision (Prec) are shown (Recall is equivalent to TPVF). 
		Hausdorff (HD) and Average Symmetric Surface Distance (ASSD) are measured in pixels.}
		\label{tab:segmentationresults}
\end{table}

\begin{table}[ht]
  \begin{center}
    \begin{tabularx}{\linewidth}{|>{\centering}l|r|}
    \hline
    Correctly diagnosed with DDH {\footnotesize (Sensitivity or TPR)} & 90.6\% \\
    \hline
    Missed DDH {\footnotesize (FNR)} & 9.4\% \\
    \hline
    \hline
    Correctly diagnosed no DDH {\footnotesize (Specificity or TNR) }& 88.2\% \\
    \hline
    Incorrectly diagnosed with DDH {\footnotesize (FPR)} & 11.8\% \\
    \hline
    \hline
    Percentage of hips correctly diagnosed {\footnotesize (Accuracy)} & 89.8\% \\
    \hline
    \end{tabularx}
  \end{center}
    \caption{Comparison of algorithm diagnosis \AFHC\ against \SGT. True Positive Rate (TPR) is the same as Sensitivity, True Negative Rate (TNR) is the same as Specificity. FPR = 1-TNR. FNR = 1-TPR.}
    \label{tab:diagnosis}
\end{table}

Tab.~\ref{tab:FHCagreement} indicates that the algorithm outperforms \GGT, which mirrors the pragmatic clinical practice. The algorithm does not overly miss-predict, but may struggle to identify the precise value in cases where DDH is borderline (Fig.~\ref{fig:scatterFHC}).

\begin{figure}[t]
\centering
\begin{minipage}[b]{.3\linewidth}
  \centerline{\includegraphics[width=\linewidth]{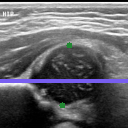}}
  \centerline{(a)}
\end{minipage}
~~~~~
\begin{minipage}[b]{.3\linewidth}
  \centerline{\includegraphics[width=\linewidth]{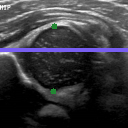}}
  \centerline{(b)}
\end{minipage}\\

\begin{minipage}[b]{.3\linewidth}
  \centerline{\includegraphics[width=\linewidth]{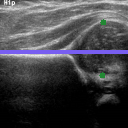}}
  \centerline{(c)}
\end{minipage}
~~~~~
\begin{minipage}[b]{.3\linewidth}
  \centerline{\includegraphics[width=\linewidth]{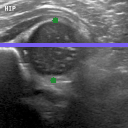}}
  \centerline{(d)}
\end{minipage}

\caption{Qualitative analysis of \AFHC\% and its corresponding diagnostic outcomes: 
(a)~True positive.
(b)~True negative.
(c)~False positive.
(d)~False negative diagnosis.
}
\label{fig:qual}
\end{figure}

The correlation statistics are already high enough for our automated tool to be widely adopted for infant hip screening. It is nevertheless essential to acknowledge that the false negative cases are not to be taken lightly. 
In Fig.~\ref{fig:qual} we illustrate the subtle differences between individual scans from the four possible categories.
Fig.~\ref{fig:qual}(a) shows a true positive example, with \AFHC=39.6\%, an error of less than 0.04\% to \SGT.
It is worth noting that, for $128{\times}128$ pixel images, a disagreement of one pixel represents a percentage difference of 0.78\% across the whole image, and approximately 1.5\% across the span of the femoral head structure.
Fig.~\ref{fig:qual}(b) shows a true negative example, with \AFHC=62.4\%, an error of less than 0.39\% to \SGT. 
Fig.~\ref{fig:qual}(c) shows the most severe of the false positives. The algorithm under-segments the lowermost component of the femoral head, with \AFHC=43.3\%, compared to a \SGT\ \FHC\ of 50.4\%.
Fig.~\ref{fig:qual}(d) shows the most severe of the false negatives. The algorithm over segments the lowermost component of the femoral head and also slightly over segments the ilium on its upper edge by including a small bump of cartilage. There is only little of the horizontal length of the ilium visible in the image; this affects the ilium edge disproportionately, and thus the algorithm calculates \AFHC=60.1\%, whereas the \SGT\ \FHC\ is 49.9\% (a figure which would have caused a clinician to keep the patient under observation).

%
%
%

\begin{table}[ht]
  \begin{center}
    \begin{tabularx}{\linewidth}{|>{\centering}X|c|c|}
    \hline
    Method 1 vs.\ Method 2& Agreement\% & Cohen's Kappa\\
    \hline
    \hline
    \AFHC\ vs.\ \SGT & {\bf 89.8\%} & {\bf 0.778}\\
    \hline
    ~~\GGT\ vs.\ \SGT & 85.7\% & 0.689\\
    \hline
    \hline
    \AFHC\ vs.\ \GGT& 87.8\% & 0.737\\
    \hline
    \end{tabularx}
  \end{center}
    \caption{Pairwise comparison of \AFHC, \SGT\ and \GGT.}
    
    \label{tab:FHCagreement}
\end{table}

\section{Conclusion and Future Work}
\label{sec:conclusion}

To our knowledge this is the first study to estimate DDH severity from UK ultrasound images using machine learning, and outperforms the existing international state of the art for automated DDH screening.
Our method makes a substantial contribution towards assistive algorithms that facilitate automating universal DDH diagnosis for every newborn.
%

In a screening setting, the 50\% decision value could be increased thus increasing sensitivity (at the cost of lowering specificity), allowing the algorithm to conservatively filter out healthy hips, maximising the focus of specialist clinical time on DDH patients and borderline cases. If clinical trials support this, the currently prohibitive workload and cost, of universal sonographic screening becomes affordable.
%

Although we have not explicitly differentiated here between dysplastic and dislocated hips, this can be easily achieved through the setting of a further decision value. 

When a full clinical dataset becomes available, we~will compare \AFHC\ against multiple clinicians, showing the automated diagnosis comes within inter-rater variability. We~will replace the segmentation CNN with a more expressive model, thereby achieving more accurate segmentations. We~expect this to improve the \AFHC\ accuracy further.

\section{Acknowledgements}
\label{sec:acknowledgments}


The authors are grateful to the \trust\ who made the anonymised data available and oversaw the ethical approval for its use for research, and to Daniel Perry and Sandeep Hemmadi who provided the \GGT.
The study was conducted retrospectively.
Each patient’s parent/carer had already agreed to  sharing  their child’s ultrasound image for this purpose.
%
%
%
%
%
The authors have no conflicts of interest to declare.


\bibliographystyle{IEEEbib}
\bibliography{2023-refs-short}

\end{document}